\documentclass[twocolumn,showpacs,showkeys,preprintnumbers,amssymb,aps,superscriptaddress,pre]{revtex4}
\usepackage{graphicx}
\usepackage{dcolumn}
\usepackage{bm}

\usepackage{epsfig}
\usepackage{color}

\begin{document}

\preprint{APS/123-QED}

\title{Vigorous thermal excitations in a double-tetrahedral chain of localized Ising spins \\ and mobile electrons mimic a temperature-driven first-order phase transition}
\author{Lucia G\'{a}lisov\'{a}}
\email{galisova.lucia@gmail.com}
\affiliation{Department of Applied Mathematics and Informatics, Faculty of Mechanical Engineering, Technical University, Letn\'{a} 9, 042 00 Ko\v{s}ice, Slovak Republic}
\author{Jozef Stre\v{c}ka}
\email{jozef.strecka@upjs.sk}
\affiliation{Department of Theoretical Physics and Astrophysics, Faculty of Science, P. J. \v{S}af\'{a}rik University, Park Angelinum 9, 040 01, Ko\v{s}ice, Slovak Republic}

\date{\today}

\begin{abstract}
A  hybrid spin-electron system defined on one-dimensional double-tetrahedral chain, in which the localized Ising spin regularly alternates with two mobile electrons delocalized over a triangular plaquette, is exactly solved with the help of generalized decoration-iteration transformation. It is shown that a macroscopic degeneracy of ferromagnetic and ferrimagnetic ground states arising from chiral degrees of freedom of the mobile electrons cannot be lifted by a magnetic field in contrast to a macroscopic degeneracy of the frustrated ground state, which appears owing to a kinetically-driven frustration of the localized Ising spins. An anomalous behavior of all basic thermodynamic quantities can be observed on account of massive thermal excitations, which mimic  a temperature-driven first-order phase transition from the non-degenerate frustrated state to the highly degenerate ferrimagnetic state at non-zero magnetic fields. A substantial difference in the respective degeneracies is responsible for an immense low-temperature peak of the specific heat and very abrupt (almost discontinuous) thermal variations of the entropy and sublattice magnetizations.
\end{abstract}


\maketitle

\section{Introduction}

Exactly solvable models are of great importance in statistical physics because they offer a valuable insight into diverse aspects of quantum, cooperative and critical phenomena \cite{Bax82,Mat93,Sut04}. It is worthwhile to remark that an exact solvability of the most famous lattice-statistical models is usually restricted to one dimension only, while the list of two- and three-dimensional rigorously solved models is much more limited \cite{Mat93}. This fact closely relates to a rather intricate nature of mathematical treatment, which must be employed in seeking an exact solution of even relatively simple interacting many-body systems \cite{Gut05}. A particularly fruitful idea for suggesting novel exactly soluble models with peculiar quantum manifestations consists in linking relatively small quantum systems through classical Ising spins. To get a closed-form exact solution for these hybrid classical-quantum models one may take advantage of generalized algebraic transformations, which establish a rigorous mapping correspondence with a simpler (fully classical) lattice-statistical model with the known exact solution \cite{Fis59,Syo72,Roj09,Str10}.

Until recently, the concept of algebraic mapping transformations has been widely applied mainly to the Ising-Heisenberg spin systems, which are composed of small clusters of quantum Heisenberg spins  coupled together through classical Ising spins only (see, e.g., Refs.~\cite{Str10,Lis11,Str12,Cis13,Lis14} and references therein). However, it has been shown later on that this conceptually simple approach is also applicable for spinless fermion models when ignoring the hopping term on particular lattice sites \cite{Roj11,Roj12}, or for hybrid spin-electron systems, where finite clusters including a few mobile electrons are mutually inter-connected through the localized Ising spins in order to form either one- \cite{Per08,Per09,Lys11,Lis13,Ana12,Nal14,Car14} or two-dimensional \cite{Str09,Tan10,Gal11,Dor14} lattice.

In the present work, we will propose and exactly solve the hybrid spin-electron system on a double-tetrahedral chain in a magnetic field. To achieve an exact solvability of this model, we will suppose that the localized Ising spins placed at nodal lattice sites regularly alternate with triangular plaquettes available to mobile electrons. It is worth mentioning that the geometry of double-tetrahedral chain was theoretically introduced by Mambrini {\it et al.} \cite{Mam99} when examining the residual entropy and spin gap in the respective Heisenberg model. Since that time, 
several other models with this lattice geometry have been discussed in literature, namely, the spinless fermion model \cite{Roj12}, the Heisenberg and Hubbard models \cite{Roj03,Bat03,Mak11,Mak12} 
and the Ising-Heisenberg model \cite{Ant09,Oha10}. A possible experimental realization of the double-tetrahedral chain is realized in the copper-based polymeric chain Cu$_3$Mo$_2$O$_9$ \cite{Has08,Kur10,Mat12}.

The outline of this paper is as follows. In Sec.~\ref{sec:2} we will describe in detail the investigated spin-electron double-tetrahedral chain and then, the most important steps of an exact mapping method will be clarified. In Sec.~\ref{sec:3} we will discuss the most interesting results for the ground state, the magnetization process and temperature dependences of basic thermodynamic quantities (magnetization, entropy, specific heat). The paper ends up with a brief summary of our findings in Sec.~\ref{sec:4}.

\begin{figure}[ht]
\begin{center}
\hspace{0.25cm}
\includegraphics[width=0.9\columnwidth]{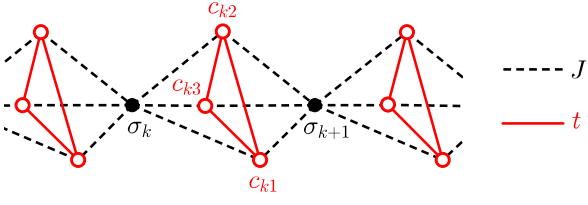}
\vspace{-0.1cm}
\caption{\small (Color online) A part of the spin-electron system on a double-tetrahedral chain. Full circles denote nodal lattice sites occupied by the localized Ising spins, while the empty circles forming triangular plaquettes are available to mobile electrons.}
\label{fig1}
\end{center}
\vspace{-0.75cm}
\end{figure}

\section{Spin-electron double-tetrahedral chain}
\label{sec:2}
Let us consider the one-dimensional double-tetrahedral chain, in which one localized Ising spin placed at nodal lattice site regularly alternates with a triangular plaquette consisting of three equivalent lattice sites available to two mobile electrons (see~Fig.~\ref{fig1}). This one-dimensional spin-electron system may alternatively be viewed as the spin-$1/2$ Ising linear chain, 
the bonds of which are decorated by triangular plaquettes available to two mobile electrons. From this perspective, the total Hamiltonian can be defined as a sum over cluster Hamiltonians ${\cal H}_k$: 
\begin{eqnarray}
\label{eq:H}
{\cal H} &=& \sum_{k = 1}^N{\cal H}_k,
\end{eqnarray}
whereas each cluster Hamiltonian ${\cal H}_k$  involves all the interaction terms connected to the mobile electrons from the $k$th triangular plaquette:
\begin{eqnarray}
\label{eq:Hk}
{\cal H}_k \!\!&=&\!\! -t \sum_{\alpha = \uparrow, \downarrow}\!(c_{k1,\alpha}^{\dag}c_{k2,\alpha} \!+ c_{k2,\alpha}^{\dag}c_{k3,\alpha} \!+ c_{k3,\alpha}^{\dag}c_{k1,\alpha} \!+ {\rm h.c.})
\nonumber\\
\!\!&+&\!\! \frac{J}{2} (\sigma_{k}^{z} + \sigma_{k+1}^{z}) \sum_{j = 1}^{3}\,(n_{kj,\uparrow} - n_{kj,\downarrow}) + U \sum_{j = 1}^{3}n_{kj,\uparrow}n_{kj,\downarrow} \nonumber\\
\!\!&-&\!\! \frac{H_{\rm I}}{2} (\sigma_{k}^{z} + \sigma_{k+1}^{z}) - \frac{H_{\rm e}}{2} \sum_{j = 1}^{3}\,(n_{kj,\uparrow} - n_{kj,\downarrow}).
\end{eqnarray}
Above, $c_{kj,\alpha}^{\dag}$ and $c_{kj,\alpha}$ represent usual fermionic creation and annihilation operators for mobile electrons from the $k$th triangular plaquette with spin $\alpha$ = $\uparrow$ or $\downarrow$, $n_{kj,\alpha} = c_{kj,\alpha}^{\dag}c_{kj,\alpha}$ is the respective number operator,  $\sigma_{k}^{z}= \pm 1/2$ labels the Ising spin placed at the $k$th nodal lattice site and $N$ denotes the total number of nodal lattice sites. The hopping parameter $t>0$ takes into account the kinetic energy of mobile electrons delocalized over triangular plaquettes, $U\geq 0$ represents the on-site Coulomb repulsion between two electrons of opposite spins occupying the same lattice site and $J$ stands for the Ising coupling between the mobile electrons and their nearest Ising neighbors. Finally, $H_{\rm I}$ and $H_{\rm e}$ are the Zeeman's terms accounting for the magnetostatic energy of the localized Ising spins and mobile electrons in a presence of the external magnetic field.

A crucial step of our calculations lies in the evaluation of the partition function for the investigated spin-electron double-tetrahedral chain. With regard to a validity of the commutation relation between different cluster Hamiltonians $[{\cal H}_k, {\cal H}_l] = 0$ ($k\neq l$), the partition function ${\cal Z}$ can be partially factorized into a product of cluster partition functions 
${\cal Z}_k$:
\begin{eqnarray}
\label{eq:Z}
{\cal Z}&=&\sum_{\{\sigma_k\}} \prod_{k=1}^N\mathrm{Tr}_k\mathrm{e}^{-\beta {\cal H}_k } = \sum_{\{\sigma_k\}} \prod_{k=1}^N{\cal Z}_k,
\end{eqnarray}
where $\beta=1/T$ is the inverse temperature (we set $k_{\rm B}=1$), the symbol $\sum_{\{\sigma_k \}}$ denotes a summation over all possible states of the localized Ising spins and the symbol $\mathrm{Tr}_k$ labels a trace over degrees of freedom of two mobile electrons from the $k$th triangular plaquette. The cluster partition function ${\cal Z}_k$ can be subsequently acquired by a diagonalization of the cluster Hamiltonian (\ref{eq:Hk}). The relevant calculation is easy to accomplish in a matrix representation of the Hilbert subspace corresponding to the cluster Hamiltonian (\ref{eq:Hk}), which is spanned over the following orthonormal basis of the electron states:
\begin{eqnarray}
|\psi_k \rangle = \{ c_{k1,\uparrow}^{\dag}c_{k2,\uparrow}^{\dag}|0 \rangle, c_{k2,\uparrow}^{\dag}c_{k3,\uparrow}^{\dag}|0 \rangle, c_{k3,\uparrow}^{\dag}c_{k1,\uparrow}^{\dag}|0 \rangle, 
\nonumber \\
              c_{k1,\downarrow}^{\dag}c_{k2,\downarrow}^{\dag}|0 \rangle, c_{k2,\downarrow}^{\dag}c_{k3,\downarrow}^{\dag}|0 \rangle, c_{k3,\downarrow}^{\dag}c_{k1,\downarrow}^{\dag}|0 \rangle, \nonumber \\
              c_{k1,\uparrow}^{\dag}c_{k1,\downarrow}^{\dag}|0 \rangle, c_{k2,\uparrow}^{\dag}c_{k2,\downarrow}^{\dag}|0 \rangle, c_{k3,\uparrow}^{\dag}c_{k3,\downarrow}^{\dag}|0 \rangle, 
\nonumber \\
              c_{k1,\uparrow}^{\dag}c_{k2,\downarrow}^{\dag}|0 \rangle, c_{k2,\uparrow}^{\dag}c_{k3,\downarrow}^{\dag}|0 \rangle, c_{k3,\uparrow}^{\dag}c_{k1,\downarrow}^{\dag}|0 \rangle, 
\nonumber \\           
              c_{k1,\downarrow}^{\dag}c_{k2,\uparrow}^{\dag}|0 \rangle, c_{k2,\downarrow}^{\dag}c_{k3,\uparrow}^{\dag}|0 \rangle, c_{k3,\downarrow}^{\dag}c_{k1,\uparrow}^{\dag}|0 \rangle \}.
\label{eq:mr}
\end{eqnarray}
($|0 \rangle$ labels the vacuum state). A straightforward diagonalization of the cluster Hamiltonian~(\ref{eq:Hk}) in the relevant Hilbert subspace gives fifteen eigenenergies:
\begin{eqnarray}
\label{eq:Ek}
{\cal E}_{1,2}&=& -\, h_{\rm I}  +\, h_{\rm e} - t, \nonumber\\
{\cal E}_{3,4}&=& -\, h_{\rm I} -\, h_{\rm e} - t, \nonumber\\
{\cal E}_{5,6}&=& -\, h_{\rm I} \pm\, h_{\rm e} +2t, \nonumber\\
{\cal E}_{7,8}&=& -\, h_{\rm I}  - t, \nonumber\\
{\cal E}_{9}  &=& -\, h_{\rm I}  + 2t, \nonumber\\
{\cal E}_{10,11}&=& -\, h_{\rm I} +\frac{1}{2}\left[t+U\!+\!\!\sqrt{(U\!-t)^2\!+8t^2}\,\right]\!,\nonumber\\
{\cal E}_{12,13}&=& -\, h_{\rm I} +\frac{1}{2}\left[t+U\!-\!\!\sqrt{(U\!-t)^2\!+8t^2}\,\right]\!,\nonumber\\
{\cal E}_{14,15}&=& -\, h_{\rm I}  -\frac{1}{2}\left[2t-U\!\pm\!\!\sqrt{(U\!+2t)^2\!+32t^2}\,\right].
\end{eqnarray}
Here, we have introduced the following notation $h_{\rm I}\! = H_{\rm I}(\sigma_{k}^{z} + \sigma_{k+1}^{z})\!/2$ and $h_{\rm e} \!= J(\sigma_{k}^{z} + \sigma_{k+1}^{z}) - H_{\rm e}$ in order to write the eigenvalues (\ref{eq:Ek}) in a more abbreviated form. The complete set of the eigenvalues (\ref{eq:Ek}) allow us to obtain the resulting expression for the cluster partition function:
\begin{eqnarray}
\label{eq:Zk}
{\cal Z}_k&=& \!\!\sum_{j=1}^{15}\mathrm{e}^{-\beta{\cal E}_{j}}\!= {\rm e}^{\beta h_{\rm I}}\Big\{\left(2{\rm e}^{\beta t}\! + {\rm e}^{-2\beta t}\right)\!\left[1 \!+\! 2\!\cosh(\beta h_{\rm e})\right] \nonumber\\
&&+\,4{\rm e}^{-\beta t/2 - \beta U/2}\cosh\left[\frac{\beta}{2}\!\sqrt{(U\!-t)^2\!+8t^2}\,\right] \nonumber\\
&&+\, 2{\rm e}^{\beta t - \beta U/2}\cosh\left[\frac{\beta}{2}\!\sqrt{(U\!+2t)^2\!+32t^2}\,\right]\Big\}.
\end{eqnarray}
It should be pointed out that the cluster partition function (\ref{eq:Zk}) still depends through the newly defined parameters $h_{\rm I}$ and $h_{\rm e}$ on the Ising spins $\sigma_{k}$ and $\sigma_{k+1}$ attached to the mobile electrons from the $k$th triangular plaquette. Next, one can perform the generalized decoration-iteration mapping transformation \cite{Fis59,Syo72,Roj09,Str10}:
\begin{eqnarray}
\label{eq:dit}
{\cal Z}_k &=& \!\!\sum_{j=1}^{15}\mathrm{e}^{-\beta{\cal E}_{j}}\!= {\rm e}^{\beta h_{\rm I}}\Big\{\left(2{\rm e}^{\beta t}\! + {\rm e}^{-2\beta t}\right)\!\left[1 \!+\! 2\!\cosh(\beta h_{\rm e})\right] \nonumber\\
&&+\,4{\rm e}^{-\beta t/2 - \beta U/2}\cosh\left[\frac{\beta}{2}\!\sqrt{(U\!-t)^2\!+8t^2}\,\right] \nonumber\\
&&+\, 2{\rm e}^{\beta t - \beta U/2}\cosh\left[\frac{\beta}{2}\!\sqrt{(U\!+2t)^2\!+32t^2}\,\right]\Big\} \nonumber\\
&=& A \exp \left[\beta J_{\rm eff} \sigma_{k}^{z} \sigma_{k+1}^{z} + \beta H_{\rm eff} (\sigma_{k}^{z} + \sigma_{k+1}^{z})/2 \right],
\end{eqnarray}
which provides an exact mapping relation between the partition function ${\cal Z}$ of the spin-electron double-tetrahedral chain and, respectively, the partition function ${\cal Z}_{\rm IC}$ of the spin-$1/2$ Ising chain with the effective nearest-neighbor coupling $J_{\rm eff}$ and the effective magnetic field $H_{\rm eff}$ after substituting Eq. (\ref{eq:dit}) into Eq. (\ref{eq:Z}):
\begin{eqnarray}
\label{eq:DIT}
{\cal Z}(\beta, J, t, U, H_{\rm I}, H_{\rm e})= A^N{\cal Z}_{\rm IC}(\beta, J_{\rm eff}, H_{\rm eff}).
\end{eqnarray}
The mapping parameters $A$, $J_{\rm eff}$ and $H_{\rm eff}$ emerging in Eq.~(\ref{eq:DIT}) can be obtained from the 'self-consistency' condition of the applied decoration-iteration transformation:
\begin{eqnarray}
\label{eq:AJeffHeff}
A&=& \sqrt[4]{\left(W_- + W\right)\left(W_+ + W\right)\left(W_0 + W\right)^2},\nonumber\\
J_{\rm eff}&=& T\ln\!\left[\frac{(W_- + W)(W_+ + W)}{(W_0 + W)^2}\right], \nonumber\\
H_{\rm eff} &=& H_{\rm I} + T\ln\left(\frac{W_- + W}{W_+ + W}\right),
\end{eqnarray}
and the functions $W_{\mp}$, $W_{0}$ and $W$ are defined as:
\begin{eqnarray}
\label{eq:W}
W_{\mp} &=& \left(2{\rm e}^{\beta t} + {\rm e}^{-2\beta t}\right)\left[1 + 2\cosh\left(\beta J\mp\beta H_{\rm e}\right)\right], \nonumber\\
W_{0} &=& \left(2{\rm e}^{\beta t} + {\rm e}^{-2\beta t}\right)\left[1 + 2\cosh\left(\beta H_{\rm e}\right)\right], \nonumber\\
W &=& 4{\rm e}^{-\beta t/2 - \beta U/2}\cosh\!\left[\frac{\beta}{2}\sqrt{(U\!-t)^2\!+8t^2}\,\right] \nonumber\\
&&+ 2{\rm e}^{\beta t - \beta U/2}\cosh\!\left[\frac{\beta}{2}\sqrt{(U\!+2t)^2\!+32t^2}\,\right]\!\!.
\end{eqnarray}
Note that the partition function of the spin-$1/2$ Ising chain in a magnetic field has exactly been calculated using the transfer-matrix method \cite{Bax82,Kra44}. From this point of view, an exact calculation of the partition function of the spin-electron double-tetrahedral chain is also formally completed.

Exact results for other thermodynamic quantities follow directly from the mapping relation~(\ref{eq:DIT}). Actually, the Gibbs free energy ${\cal G}$ of the spin-electron double-tetrahedral chain takes the form:
\begin{eqnarray}
\label{eq:G}
{\cal G}=-T\ln{\cal Z}_{\rm IC} - NT\ln A,
\end{eqnarray}
which can be further used for the calculation of the entropy $S$ and the specific heat $C$:
\begin{eqnarray}
\label{eq:SC}
S=-\left(\frac{\partial {\cal G}}{\partial T}\right)_H, \quad C=-T\left(\frac{\partial^2 {\cal G}}{\partial T^2}\right)_H,
\end{eqnarray}
as well as the sublattice magnetizations $m_{\rm I}$ and $m_{\rm e}$ normalized per one localized Ising spin and mobile electron, respectively:
\begin{eqnarray}
\label{eq:mIme}
m_{\rm I}=-\frac{1}{N}\left(\frac{\partial {\cal G}}{\partial H_{\rm I}}\right), \quad m_{\rm e}=-\frac{1}{2N}\left(\frac{\partial {\cal G}}{\partial H_{\rm e}}\right).
\end{eqnarray}
In view of this notation, the total magnetization normalized per one magnetic particle of the spin-electron double-tetrahedral chain can be expressed as:
\begin{eqnarray}
\label{eq:m}
m = \frac{1}{3}\left(m_{\rm I} + 2m_{\rm e}\right).
\end{eqnarray}

\section{Results and discussion}
\label{sec:3}
In this section, we will proceed to a discussion of the most interesting results for the investigated spin-electron double-tetrahedral chain by considering the particular case with the antiferromagnetic interaction $J>0$ between the localized Ising spins and mobile electrons. To reduce the total number of free interaction parameters, we will assume equal magnetic fields acting on the Ising spins and mobile electrons, i.e. $H_{\rm I}= H_{\rm e} \equiv H \geq 0$.

\subsection{Ground state}
\label{subsec:GS}
To get the ground state of the investigated spin-electron model, it is sufficient to find the lowest-energy eigenstate of the cluster Hamiltonian (\ref{eq:Hk}) that can be simply extended to the whole double-tetrahedral chain due to the commuting character of the cluster Hamiltonians. The lowest-energy eigenstate can be obtained by inspection from the full spectrum of eigenvalues (\ref{eq:Ek}) of the cluster Hamiltonian (\ref{eq:Hk}) after taking into account all four available states of two nodal Ising spins $\sigma_k$ and $\sigma_{k+1}$ involved therein. In this way, one finds three different macroscopically degenerate ground states: the ferromagnetic (FM) state, the ferrimagnetic (FRI) state and the frustrated (FRU) state, which are unambiguously characterized by the following eigenvectors and energies:
\begin{eqnarray}
\label{eq:FM}
|{\rm FM}\rangle \!\!&=&\!\! \prod_{k=1}^N |\!\uparrow\rangle_{\sigma_k} \otimes \left\{ \begin{tabular}{c}
                                                                        $|\phi_{\uparrow}^{+}\rangle_k$ \\
                                                                        $|\phi_{\uparrow}^{-}\rangle_k$
                                                                       \end{tabular} \right., \nonumber\\
{\cal E}_{\rm FM} \!\!&=&\!\! \frac{N}{2}\left(2J - 3H - 2t\right); \\
\label{eq:FRI1}
|{\rm FRI}\rangle \!\!&=&\!\! \prod_{k=1}^N |\!\downarrow\rangle_{\sigma_k} \otimes \left\{ \begin{tabular}{c}
                                                                        $|\phi_{\uparrow}^{+}\rangle_k$ \\
                                                                        $|\phi_{\uparrow}^{-}\rangle_k$
                                                                       \end{tabular} \right., \nonumber\\
{\cal E}_{\rm FRI} \!\!&=&\!\! -\frac{N}{2}\left(2J + H +2t\right); \\
\label{eq:FRI2}
|{\rm FRU}\rangle \!\!&=&\!\!  \Bigg{ \{ } \begin{tabular}{c} ${\displaystyle{\prod_{k=1}^N}}$ 
                                                                \begin{tabular}{c}
                                                                        $|\!\uparrow\rangle_{\sigma_k}$ \\
                                                                        $|\!\downarrow\rangle_{\sigma_k}$
                                                                       \end{tabular} \Big\} $\otimes$ $|\phi_{0}\rangle_k$,  $\qquad$ $H = 0$ \nonumber \\  
                                                            ${\displaystyle{\prod_{k=1}^N}}$  $|\!\uparrow\rangle_{\sigma_k}$ $\otimes$ $|\phi_{0}\rangle_k$,  $\qquad$ $H > 0$ \nonumber \\        
                                                     \end{tabular}   \nonumber\\                                                                                             
{\cal E}_{\rm FRU} \!\!&=&\!\! \frac{N}{2}\left(U-H-2t - \!\sqrt{(U+2t)^2+32t^2}\,\right)\!\!.
\end{eqnarray}
In above, the product runs over all primitive unit cells, the state vector $|\!\uparrow\rangle_{\sigma_k}$ ($|\!\downarrow\rangle_{\sigma_k}$) determines up (down) state of the $k$th localized Ising spin $\sigma_k^z=1/2$ $(\sigma_k^z=-1/2)$. The state vectors $|\phi^{+}_{\uparrow}\rangle_k$ and $|\phi^{-}_{\uparrow}\rangle_k$ emerging in Eqs.~(\ref{eq:FM}) and~(\ref{eq:FRI1}) label two eigenstates of the mobile electrons with a positive and negative chirality:
\begin{eqnarray}
|\phi^{+}_{\uparrow}\rangle_k  &=& \frac{1}{\sqrt{3}}\,
(c_{k1,\uparrow}^{\dag}c_{k2,\uparrow}^{\dag}\! +\omega c_{k2,\uparrow}^{\dag}c_{k3,\uparrow}^{\dag}\! +\omega^{2} c_{k3,\uparrow}^{\dag}c_{k1,\uparrow}^{\dag})|0\rangle, \nonumber\\
|\phi^{-}_{{\uparrow}}\rangle_k  &=& \frac{1}{\sqrt{3}}\,
(c_{k1,\uparrow}^{\dag}c_{k2,\uparrow}^{\dag}\! +\omega^{2} c_{k2,\uparrow}^{\dag}c_{k3,\uparrow}^{\dag}\! +\omega c_{k3,\uparrow}^{\dag}c_{k1,\uparrow}^{\dag})|0\rangle,\nonumber \\
\omega &=& {\rm e}^{2\pi {\rm i}/3} \,\, \qquad({\rm i} = \sqrt{-1}), 
\label{eq:electron1}
\end{eqnarray}
while the last eigenstate of the mobile electrons $|\phi_{0}\rangle_k$ appearing in Eq.~(\ref{eq:FRI2}) refers to a non-chiral state with a zero current:
\begin{eqnarray}
|\phi_{0}\rangle_k &=& \frac{1}{\sqrt{6}}\,
\big[\sin\varphi\,(c_{k1,\uparrow}^{\dag}c_{k2,\downarrow}^{\dag}\!+ c_{k2,\uparrow}^{\dag}c_{k3,\downarrow}^{\dag} \!+ c_{k3,\uparrow}^{\dag}c_{k1,\downarrow}^{\dag} \nonumber\\
& &-\, c_{k1,\downarrow}^{\dag}c_{k2,\uparrow}^{\dag}\! - c_{k2,\downarrow}^{\dag}c_{k3,\uparrow}^{\dag} \!
 - c_{k3,\downarrow}^{\dag}c_{k1,\uparrow}^{\dag}) \nonumber\\
& & +  \sqrt{2}\cos\varphi\sum_{j=1}^{3}c_{kj,\uparrow}^{\dag}c_{kj,\downarrow}^{\dag}\big]|0\rangle.
\label{eq:electron2}
\end{eqnarray}
The mixing angle $\varphi$ determining a quantum entanglement of the relevant electron states within the eigenstate (\ref{eq:electron2}) depends on a mutual competition between the hopping term and Coulomb term through the relation ${\rm tan}\,\varphi =\!\frac{\sqrt{2}}{8t}(U+2t+\sqrt{(U+2t)^2+32t^2})$. 

It can be easily understood from Eqs.~(\ref{eq:FM}) and~(\ref{eq:FRI1}) that the common feature of FM and FRI ground states is a quantum entanglement of three ferromagnetic states $c_{k1,\uparrow}^{\dag}c_{k2,\uparrow}^{\dag}|0\rangle$, $c_{k2,\uparrow}^{\dag}c_{k3,\uparrow}^{\dag}|0\rangle$, $c_{k3,\uparrow}^{\dag}c_{k1,\uparrow}^{\dag}|0\rangle$ of the mobile electrons and hence, both ground states differ from each other just by the respective spin arrangement of the localized Ising spins. While the Ising spins are aligned into a direction of the external magnetic field within the FM ground state appearing at high enough magnetic fields, they point in opposite direction in the FRI ground state with regard to the antiferromagnetic coupling with the mobile electrons. It should be stressed, moreover, that the lowest-energy eigenstate (\ref{eq:electron1}) of the mobile electrons is two-fold degenerate at each triangular plaquette due to two possible values of the scalar chirality, which consequently leads to a substantial residual entropy $S/3N = \ln2^{1/3}$ of the FM and FRI ground states.

However, the most spectacular ground state is realized in the FRU phase, in which the mobile electrons underlie a quantum superposition of six intrinsic antiferromagnetic states  $c_{k1,\uparrow}^{\dag}c_{k2,\downarrow}^{\dag}|0\rangle$, $c_{k2,\uparrow}^{\dag}c_{k3,\downarrow}^{\dag}|0\rangle$,  $c_{k3,\uparrow}^{\dag}c_{k1,\downarrow}^{\dag}|0\rangle$,  $c_{k1,\downarrow}^{\dag}c_{k2,\uparrow}^{\dag}|0\rangle$, $c_{k2,\downarrow}^{\dag}c_{k3,\uparrow}^{\dag}|0\rangle$, $c_{k3,\downarrow}^{\dag}c_{k1,\uparrow}^{\dag}|0\rangle$ and three non-magnetic ionic states $c_{kj,\uparrow}^{\dag}c_{kj,\downarrow}^{\dag}|0\rangle$ ($j=1,2,3$). The nature of the FRU ground state consists in a kinetically-driven spin frustration, the origin of which is quite similar to that reported previously by Pereira \textit{et al}. for the analogous spin-electron diamond chain \cite{Per08,Per09}. As a matter of fact, the localized Ising spins are at a zero magnetic field completely free to flip in arbitrary direction owing to the kinetically-driven spin frustration caused by the antiferromagnetic alignment of the mobile electrons, whereas arbitrary but non-zero magnetic field tends to align them into the external-field direction. The magnetic field thus lifts the macroscopic degeneracy of the FRU ground state (and the associated residual entropy $S/3N = \ln2^{1/3}$) in contrast with two previously discussed FM and FRI ground states, where the macroscopic degeneracy relates to chiral degrees of freedom of the mobile electrons rather than to a kinetically-driven spin frustration of the localized Ising spins. 

\begin{figure}[t]
\centering
\hspace{0.25cm}
\includegraphics[width = 0.9\columnwidth]{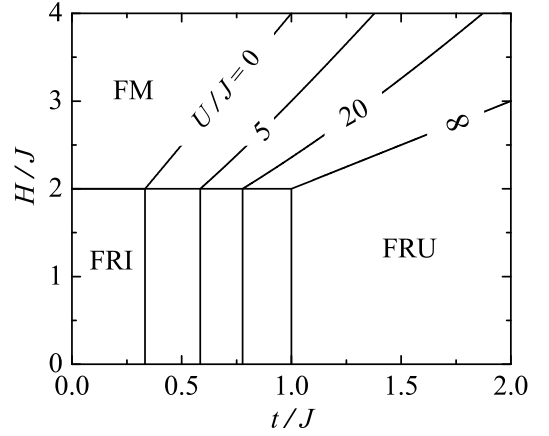}
\vspace{-0.5cm}
\caption{\small The ground-state phase diagram in the $t/J-H/J$ plane for the spin-electron double-tetrahedral chain with the antiferromagnetic coupling $J>0$  upon varying a relative strength of the Coulomb term $U/J = 0$, $5$, $20$, $\infty$.}
\label{fig2}
\end{figure}

The ground-state phase diagram in $t/J-H/J$ plane involving all three possible ground states is depicted on Fig.~\ref{fig2} for several values of the Coulomb term $U/J$. Evidently, the FRI phase becomes a ground state at low enough magnetic fields whenever the influence of the antiferromagnetic Ising interaction $J$ overwhelms the effect of the hopping term $t$, i.e. whenever the hopping term is smaller $t<t_{\rm b}$ than the boundary value  
\begin{eqnarray}
t_{\rm b} &=& -\frac{U}{18} + \frac{1}{18}\sqrt{(U + 6J)^2+ 24UJ}.
\label{eq:FRI1FRI2}
\end{eqnarray}
If the reverse condition $t>t_{\rm b}$ holds, then, the FRU phase is preferred as a ground state at sufficiently low magnetic fields due to a predominant influence of the kinetically-driven spin frustration. Of course, the investigated spin-electron double-tetrahedral chain undergoes at high enough magnetic fields a field-induced phase transition towards the FM ground state with all nodal Ising spins and mobile electrons fully polarized to the external-field direction. Analytic expressions of the relevant first-order phase transitions read:
\begin{eqnarray}
\label{eq:FRI1FM}
{\rm FRI} \!- {\rm FM}\!:H \!\!&=&\!\! 2J,\\
\label{eq:FRI2FM}
{\rm FRU} \!- {\rm FM}\!:H \!\!&=&\!\! J - \frac{U}{2}+\frac{1}{2}\!\sqrt{(U + 2t)^2\!+32t^2}.
\end{eqnarray}
It is worthwhile to remark that the macroscopic degeneracy $S/3N = \ln4^{1/3}\approx0.4621$ at the FRI--FM phase boundary is greater than the macroscopic degeneracy $S/3N = \ln3^{1/3}\approx0.3662$ at the FRU--FM phase boundary. 

To complete our ground-state analysis, let us make a few comments on a special limit of infinitely strong Coulomb repulsion $U/t \to \infty$ when a mutual effect of the hopping and Coulomb term should be equivalent to the antiferromagnetic Heisenberg coupling as it can be proved within the second-order perturbation theory \cite{Faz99}. The ground-state phase diagram of the investigated spin-electron double-tetrahedral chain with two electrons per triangular plaquette indeed becomes identical in the $U/t \to \infty$ limit with the ground-state phase diagram of the spin-1/2 Ising-Heisenberg diamond chain (compare Fig. \ref{fig2} with Fig. 2b in Ref. \cite{Can06}). However, there is a fundamental difference in a character of the relevant ground states: all three ground states of the spin-electron double-tetrahedral chain exhibit a remarkable quantum entanglement in contrast to a classical nature of spin arrangements of two ground states (FRI and SPP) of the spin-1/2 Ising-Heisenberg diamond chain \cite{Can06}. Besides, it is quite obvious that essential magnetic features of the model under investigation do not qualitatively change with the Coulomb term and hence, our further analysis will be restricted to the particular case with the fixed value of the on-site Coulomb repulsion $U/J = 5$.

\begin{figure}[t]
\centering
\hspace{0.25cm}
\includegraphics[width = 0.9\columnwidth]{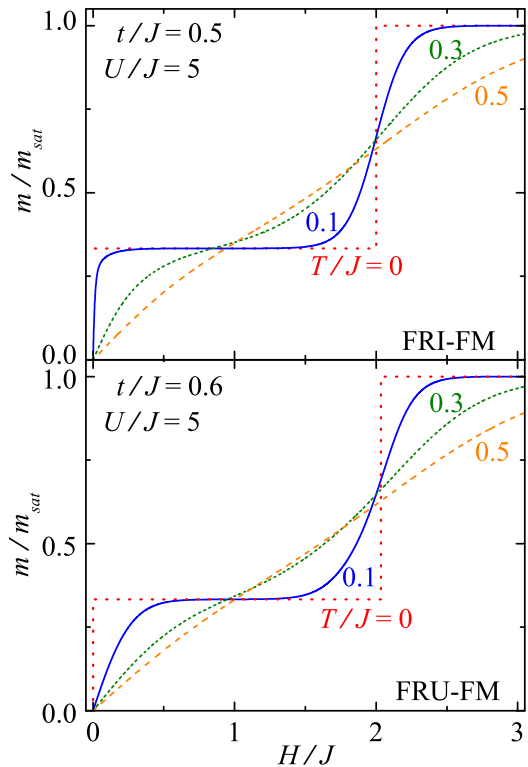}
\vspace{-0.5cm}
\caption{\small (Color online) The total magnetization normalized with respect to its saturation value as a function of the magnetic field at a few temperatures, the fixed value of the Coulomb term $U/J = 5$ and two different values of the hopping term $t/J=0.5$ (upper panel), $t/J=0.6$ (lower panel).}
\label{fig3}
\end{figure}

\subsection{Magnetization process}
\label{subsec:MAG}
The spin-electron double-tetrahedral chain bears a close relation to the spin-1/2 Ising-Heisenberg diamond chain \cite{Can06} as far as the behavior of magnetic quantities is concerned. To illustrate this point, the total magnetization is plotted in Fig. \ref{fig3} against the magnetic field for the fixed value of the Coulomb term, two values of the hopping term and a few temperatures. As one can see, the displayed magnetization curves are quite similar to that of the spin-1/2 Ising-Heisenberg diamond chain (cf. with Fig. 3 of Ref. \cite{Can06}). In fact, the intermediate one-third plateau can always be detected in low-temperature magnetization curves irrespective of whether the FRI or FRU phase is realized as the ground state before the magnetization reaches saturation at sufficiently high magnetic fields. In addition, the zero-temperature magnetization curve starts from non-zero value in the asymptotic limit of vanishing external field if the ground state is formed by the FRI phase (the upper panel in Fig. \ref{fig3}), while it starts from zero asymptotic limit on assumption that the FRU phase constitutes the ground state (the lower panel in Fig. \ref{fig3}). Both aforementioned features have been already found and discussed in depth in our previous work concerned with the spin-1/2 Ising-Heisenberg diamond chain \cite{Can06}. In agreement with common expectations, the one-third plateau as well as a steep increase in the magnetization observable near zero and saturation fields are gradually smoothened upon increasing temperature.

\begin{figure}[t]
\centering
\hspace{0.25cm}
\includegraphics[width = 0.9\columnwidth]{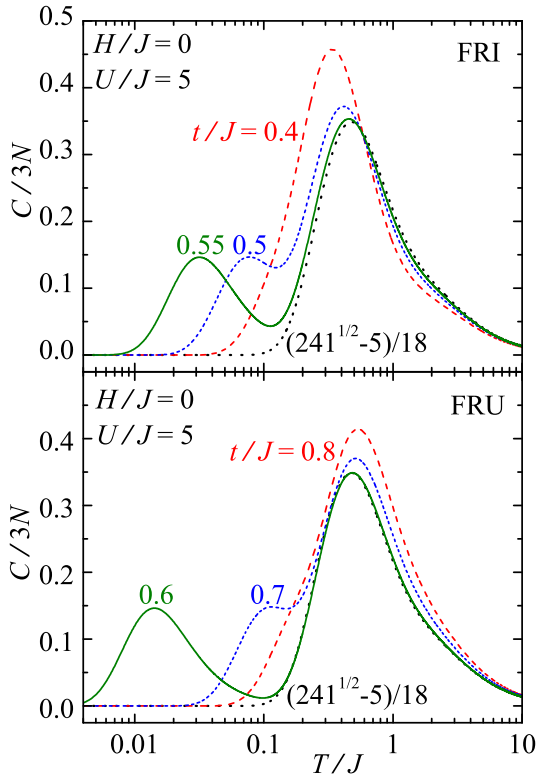}
\vspace{-0.5cm}
\caption{\small (Color online) Temperature variations of the zero-field specific heat for the special value of the Coulomb term $U/J = 5$ and a few different values of the hopping term. The upper (lower) panel shows thermal dependences when the hopping term is smaller (greater) than the boundary value $t_{\rm b}/J = (\sqrt{241}-5)/18$. The lowest dotted curve illustrates thermal dependence exactly at the FRI--FRU phase boundary.}
\label{fig4}
\end{figure}

\subsection{Low-temperature thermodynamics}
\label{subsec:CS}
Let us examine in detail temperature variations of basic thermodynamic quantities such as the specific heat and entropy. Fig. \ref{fig4} shows temperature dependences of the zero-field specific heat for a few different values of the kinetic term $t/J$. According to Eq. (\ref{eq:FRI1FRI2}), a relative size of the hopping term as compared with the boundary value $t_{\rm b}/J = (\sqrt{241}-5)/18 \approx 0.5847$ is conclusive whether the FRI or FRU phase represents the actual ground state when considering the special value of Coulomb repulsion $U/J = 5$. The upper (lower) panel thus demonstrates temperature variations of the zero-field specific heat, which are quite typical for the particular case with the FRI (FRU) ground state. It is quite apparent from Fig. \ref{fig4} that  temperature dependences of the zero-field specific heat generally exhibit one broad maximum in a high-temperature region regardless of whether the FRI or FRU phase constitutes the ground state. Moreover, there also may appear one additional Schottky-type maximum at a lower temperature if the hopping term $t/J$ is selected sufficiently close to the FRI--FRU phase boundary. A position of the low-temperature peak moves towards zero temperature as the kinetic term approaches the boundary value $t_{\rm b}/J$ given by Eq. (\ref{eq:FRI1FRI2}) at which it completely disappears (see dotted curve in Fig. \ref{fig4}). It can be also seen from Fig. \ref{fig4} that the low-temperature Schottky-type peak gradually merges with a round high-temperature maximum when the size of hopping term varies further apart from the FRI--FRU phase boundary.

\begin{figure}[t]
\centering
\hspace{0.25cm}
\includegraphics[width = 0.9\columnwidth]{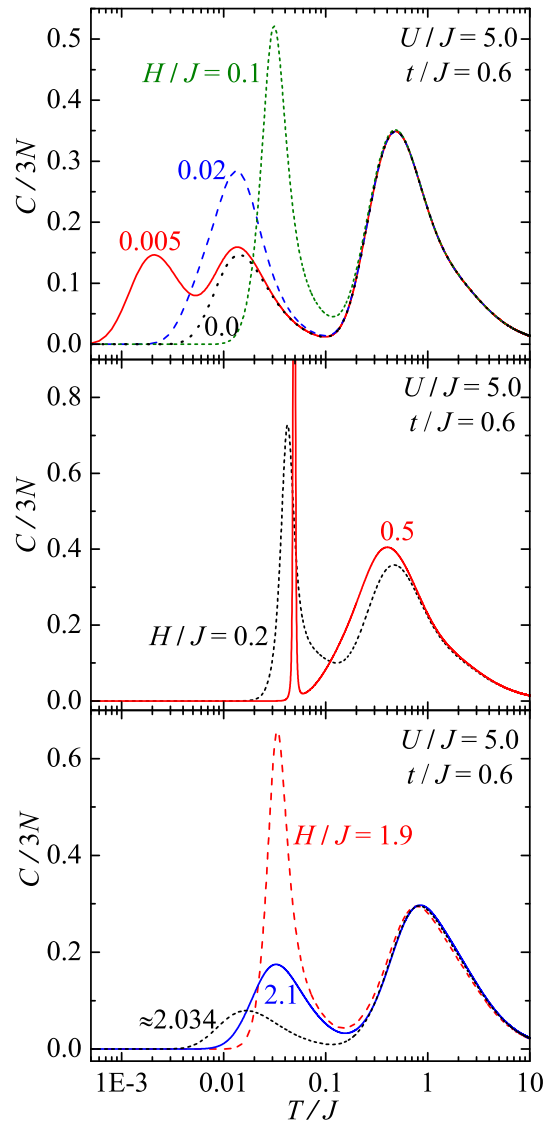}
\vspace{-0.5cm}
\caption{\small (Color online) Temperature dependences of the specific heat for the special value of the Coulomb term $U/J=5$, the hopping term $t/J=0.6$ 
and a few different values of the external magnetic field $H/J$. The lowest dotted curve in the upper panel shows the relevant zero-field dependence, 
while the lowest short-dashed curve in the lower panel shows the relevant dependence at the saturation field.}
\label{fig5}
\end{figure}

Furthermore, let us turn our attention to the effect of external magnetic field on temperature variations of the specific heat. A few typical dependences are depicted in Fig. \ref{fig5} for the particular case with the fixed values of Coulomb and hopping terms, which drive the model system to the FRU ground state in a vicinity of the FRI--FRU phase boundary. Under this condition, a relatively small applied magnetic field is responsible for the appearance of a remarkable triple-peak dependence of the specific heat, whereas the maximum found at the lowest temperature can be ascribed to the Zeeman's splitting of energy levels of the frustrated Ising spins (see the curve $H/J=0.005$ in the upper panel of Fig. \ref{fig5}). In accordance with this statement, the maximum gradually shifts towards higher temperatures with increasing magnetic field until it merges with the intermediate maximum, which is also present in the relevant zero-field dependence (dotted curve in the upper panel). The intermediate maximum appears due to a thermal excitation from the FRU ground state towards the low-lying FRI excited state once the hopping term is selected slightly above the boundary value (\ref{eq:FRI1FRI2}) of the FRI--FRU phase boundary. However, the most surprising finding is a rather abrupt rise of this low-temperature peak at moderate values of the magnetic field (see the central panel of Fig. \ref{fig5}), the origin of which will be examined in the following. Although the height of the low-temperature peak is greatest around the moderate magnetic fields $H/J \approx 1$, the low-temperature peak persists up the magnetic fields slightly above the saturation field and it does not disappear neither at the saturation field (see the short-dashed curve in the lower panel of Fig. \ref{fig5}). 

\begin{figure}[t]
\centering
\hspace{0.25cm}
\includegraphics[width=0.9\columnwidth]{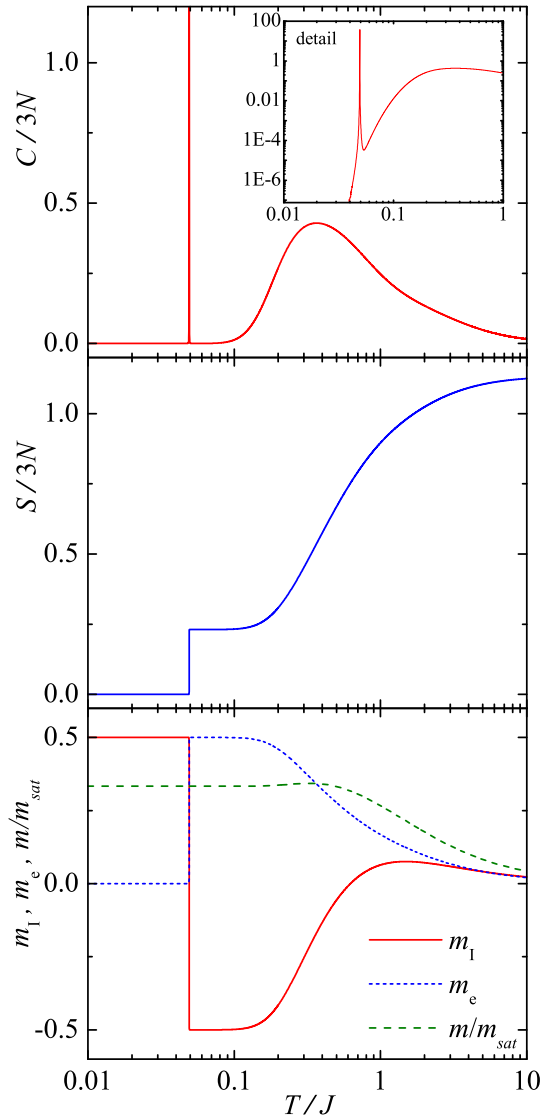}
\vspace{-0.5cm}
\caption{\small (Color online) Temperature dependences of the specific heat (upper panel), entropy (central panel) and sublattice magnetizations (lower panel) for the fixed values of the Coulomb term $U/J=5$, the hopping term $t/J=0.6$ and the magnetic field $H/J=1$. The inset shows a detailed plot of the specific heat in a low-temperature region within a log-log scale.}
\label{fig6}
\end{figure}

Let us provide a comprehensive understanding of the origin of the sizable low-temperature peak, which occurs in a thermal dependence of the specific heat at moderate values of the magnetic field. To clarify this issue, we depict in Fig. \ref{fig6} temperature variations of the specific heat, entropy and sublattice magnetizations for the set of parameters $U/J=5$, $t/J=0.6$ and $H/J=1$ for which this peculiar phenomenon is especially pronounced. It is quite clear from the inset of Fig. \ref{fig6} that the specific heat shows a very sharp peak in a relatively narrow temperature range, which could easily be confused either with the $\lambda$-type divergence accompanying a second-order phase transition or the anomalous peak accompanying a first-order phase transition. Although the temperature dependence of the total magnetization does not exhibit any striking feature, both sublattice magnetizations pertinent to the localized Ising spins and mobile electrons exhibit very abrupt thermal variations (almost discontinuous jumps) from the values typical for the FRU phase to the values typical for the FRI phase (see the lower panel in Fig. \ref{fig6}). This result affords a convincing evidence that the anomalous peak in the specific heat can be attributed to vigorous thermal excitations from the FRU ground state to the FRI excited state. A rather steep increase observed in the relevant temperature dependence of entropy (the central panel of Fig. \ref{fig6}) is also in accordance with this statement, because almost discontinuous jump between zero and $\ln 2^{1/3}$ is consistent with a zero-point entropy of the FRU phase and the residual entropy of the FRI phase at non-zero magnetic fields. 

\begin{figure}[t]
\centering
\vspace{-0.05cm}
\includegraphics[width = 0.9\columnwidth]{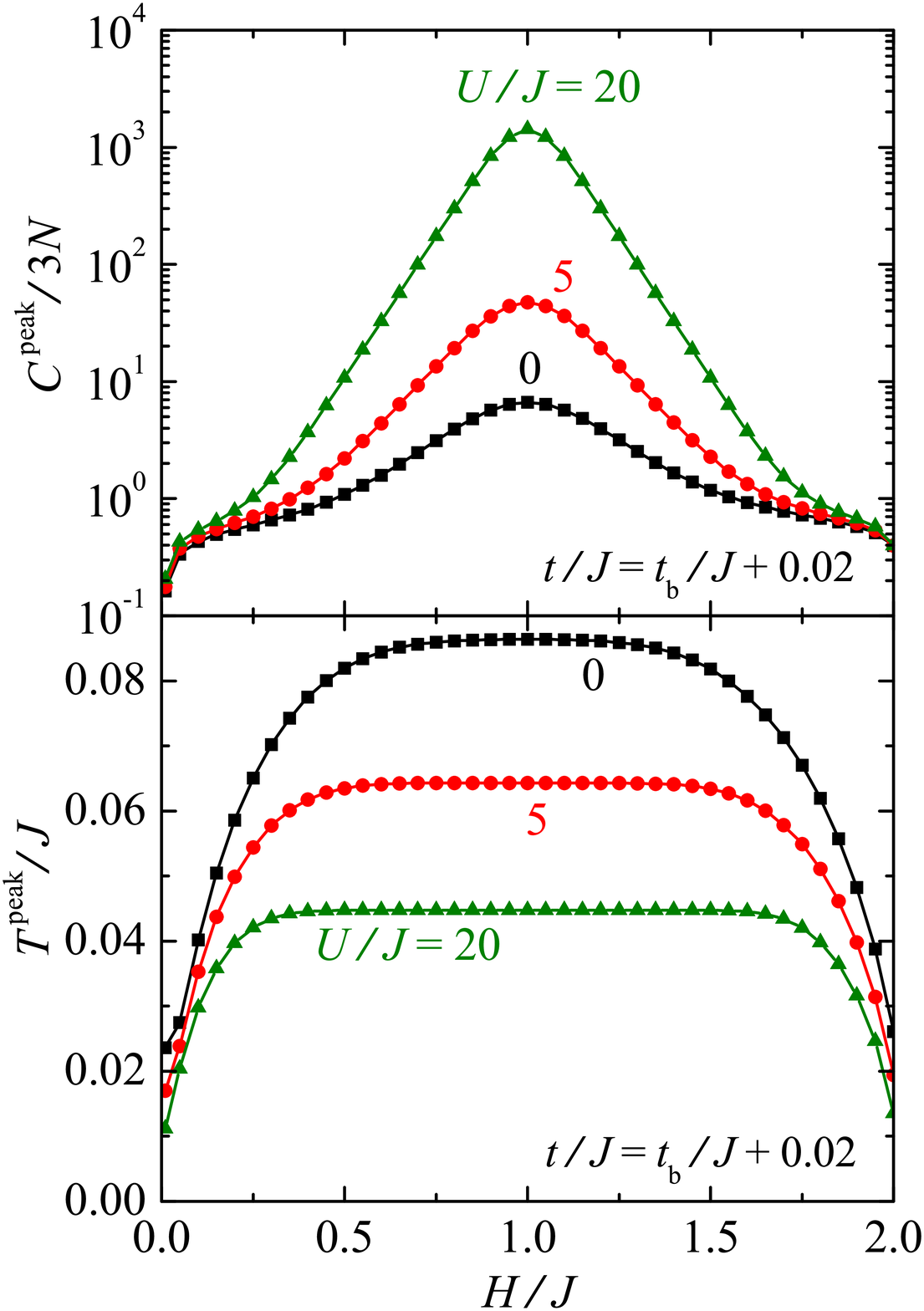}
\vspace{-0.5cm}
\caption{\small (Color online) The height (upper panel) and the position (lower panel) of the low-temperature peak of the specific heat, which occurs due to massive thermal excitations from the non-degenerate FRU ground state towards the highly degenerate FRI ground state when the hopping term is selected slightly above the FRI--FRU phase boundary $t/J = t_{\rm b}/J + 0.02$ and the Coulomb term varies $U/J = 0$, $5$, $20$.}
\label{fig7}
\end{figure}

It could be concluded that a substantial difference between degeneracies of the FRU and FRI states is responsible for vigorous thermal excitations manifested as the anomalous peak of the specific heat, which mimic a temperature-induced first-order phase transition between the FRU and FRI states. In the consequence of that, one may estimate a pseudo-critical temperature corresponding to the massive thermal excitations from the FRU phase to the FRI phase from the equality of Gibbs free energies:
\begin{eqnarray}
T_{\rm pc} = \frac{\sqrt{(U+2t)^2+32t^2}-U-2J}{\ln 4}.
\label{eq:ptc}
\end{eqnarray}  
Even if a temperature change of the enthalpy and entropy has been completely neglected by a derivation of the pseudo-critical temperature, the formula (\ref{eq:ptc}) provides a very rather accurate estimate of the peak position provided that the system is sufficiently close to the FRU--FRI phase boundary, i.e. vigorous thermal excitations between both states occur near zero temperature. Fig. \ref{fig7} illustrates changes in the height and position of the low-temperature maximum arising when the hopping term $t/J = t_{\rm b}/J + 0.02$ is selected slightly above the FRI--FRU phase boundary for a few specific values of the Coulomb term $U/J = 0$, $5$ and $20$. In agreement with previous argumentation, Fig. \ref{fig7} confirms our statement that the low-temperature peak shows the greatest height around the moderate field $H/J \approx 1$. Moreover, it can be also seen from the upper panel in Fig. \ref{fig7} that the peak height generally increases with increasing the Coulomb term. This result would suggests that the Coulomb term supports temperature-induced excitations between the FRU and FRI states. On the other hand, the peak position remains unchanged around $H/J \approx 1$ with an accuracy up to three decimal places, whereas it shifts towards lower temperatures as the Coulomb term increases (see the lower panel in Fig. \ref{fig7}). 

\subsection{Enhanced magnetocaloric effect}
\label{subsec:MCE}

\begin{figure}[ht]
\centering
\vspace{-0.1cm}
\includegraphics[width = 1.0\columnwidth]{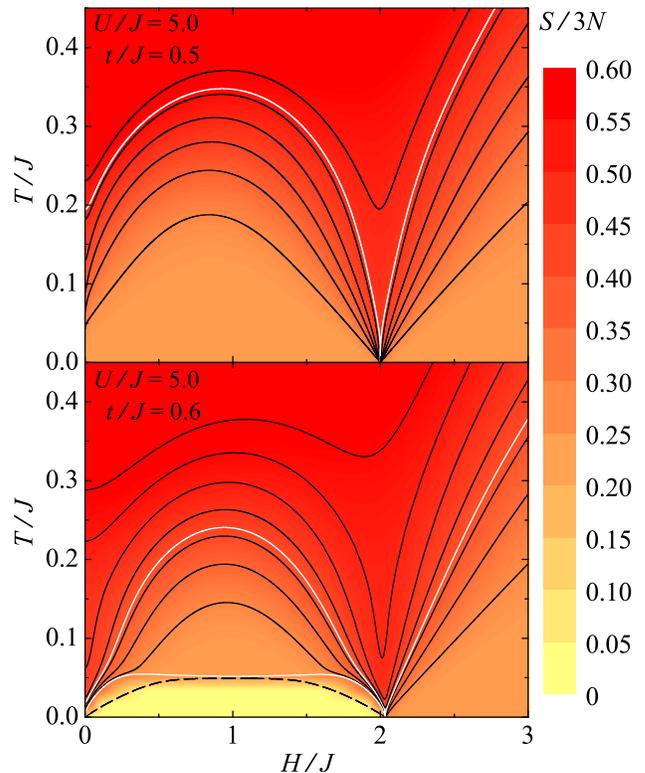}
\vspace{-1.0cm}
\caption{\small (Color online) A density plot of the entropy as a function of the magnetic field and temperature by assuming the constant value of the Coulomb term $U/J = 5$ and two different values of the hopping term $t/J=0.5$ (upper panel) and $t/J=0.6$ (lower panel). The displayed curves correspond to isentropy lines, namely, $S/3N=0.25, 0.3, $\ldots$, 0.55$ (black solid curves), $S/3N=\ln 4^{1/3}$ (white solid curve in the upper panel), $S/3N=0.001$ (black broken curve), and $S/3N=\ln 2^{1/3}, \ln 3^{1/3}$ (white solid curves in the lower panel).}
\label{fig8}
\end{figure}

Last but not least, we turn to a discussion of the magnetocaloric effect in its classical interpretation as an adiabatic change of temperature achieved upon varying the external magnetic field. For this purpose, the density plot of the entropy is depicted in Fig. \ref{fig8} as a function of the magnetic field and temperature for two different magnetization scenarios discussed previously. Isentropic changes of temperature upon varying the magnetic field can be identified in Fig. \ref{fig8} as contours of constant entropy displayed by solid lines. The adiabatic demagnetization related to the field-induced phase transition between the FRI and FM phase can be analyzed from dependences shown in the upper panel of Fig. \ref{fig8}. It is quite evident that the model under investigation exhibits an enhanced magnetocaloric effect in a vicinity of the relevant field-induced transition whenever the entropy is set sufficiently close to the value $S/3N = \ln 4^{1/3}\approx 0.4621$, 
but temperature finally tends towards a finite value as the magnetic field vanishes. On the other hand, the enhanced magnetocaloric effect during the adiabatic demagnetization can be found at zero field as well as the relevant critical field for another magnetization scenario pertinent to the field-induced phase transition between the FRU and FM phase (see the lower panel of Fig. \ref{fig7}).  Under this circumstance, the most abrupt drop in temperature is achieved under the adiabatic condition if the entropy is set sufficiently close to the values $S/3N = \ln 2^{1/3}\approx0.231$ 
and $S/3N = \ln 3^{1/3}\approx0.3662$, respectively. The adiabatic change of temperature in response to the variation in a magnetic field is unusually striking for the isentropy line $S/3N = \ln 2^{1/3}\approx0.231$ (the lower white curve in the lower panel of Fig. \ref{fig8}), for which temperature remains nearly constant in a relatively wide range of the magnetic fields $H/J \in (1/2,3/2)$. This intriguing feature can be attributed to the anomalous behavior of specific heat discussed previously in Sec. \ref{subsec:CS}, because the adiabatic change of temperature with the magnetic field is inversely proportional to the specific heat and it is therefore negligible due to a sizable value of the specific heat.    

\section{Conclusions}
\label{sec:4}

The present work deals with magnetic properties of a one-dimensional double-tetrahedral chain of localized Ising spins and mobile electrons, which can be exactly treated through the generalized decoration-iteration transformation establishing a rigorous mapping correspondence with a simple spin-1/2 Ising chain with the effective nearest-neighbor interaction and effective magnetic field. Our exact calculation have allowed us to examine in detail the ground-state phase diagram, magnetization process, magnetocaloric effect, entropy and specific heat. Although the investigated spin-electron model on a double-tetrahedral chain resembles to a certain extent some magnetic features of the spin-1/2 Ising-Heisenberg diamond chain (e.g., an intermediate one-third plateau in a low-temperature magnetization curve, enhanced magnetocaloric effect during the adiabatic demagnetization and temperature variations of the specific heat with one, two or three separate peaks) \cite{Can06}, it also displays a lot of other remarkable features not reported in the literature hitherto. 

In particular, we have found three different ground states with an interesting quantum entanglement between states of the mobile electrons and a high macroscopic degeneracy. The ferromagnetic and ferrimagnetic ground states are macroscopically degenerate due to chiral degrees of freedom of the mobile electrons, while the frustrated state displays a macroscopic degeneracy owing to a kinetically-driven frustration of the localized Ising spins. It has been evidenced that the residual entropy due to the kinetically-driven spin frustration can be thoroughly lifted by the magnetic field unlike the residual entropy connected to chiral degrees of freedom of the mobile electrons. However, the most spectacular finding concerns with the anomalous behavior of all basic thermodynamic quantities if the hopping and Coulomb terms drive the system to the frustrated ground state in a close vicinity of the phase boundary with the ferrimagnetic state. A substantial difference in the respective ground-state degeneracies is responsible for an immense low-temperature peak of the specific heat and very abrupt (almost discontinuous) thermal variations of the entropy and sublattice magnetizations. The caution before interpreting all aforementioned features as typical manifestations of a phase transition should be accordingly made \cite{Man08,Man13}, because our exactly solved spin-electron chain serves in evidence that all those outstanding features can originate from vigorous thermal excitations between a non-degenerate ground state and a highly degenerate (low-lying) excited state due to a high entropy gain.

\begin{acknowledgments}
This work was financially supported by the grant of the Slovak Research and Development Agency under the contract \mbox{No. APVV-0097-12} and by the ERDF EU (European Union European regional development
fond) grant provided under the contract No. ITMS26220120005 (activity 3.2). J.S. would like to thank Dr. Taras Verkholyak for his helpful comments and encouraging discussions concerning with the topic of this work.
\end{acknowledgments}


\begin{thebibliography}{100}
\bibitem{Bax82} R.J. Baxter, Exactly Solved Models in Statistical Mechanics (Academic Press, New York, 1982).
\bibitem{Mat93} D.C. Mattis, The Many-Body Problem: An Encyclopedia of Exactly Solved Models in One Dimension (World Scientific, Singapore, 1993).
\bibitem{Sut04} B. Sutherland, Beautiful Models: 70 Years of Exactly Solved Quantum Many-Body Problems (World Scientific, Singapore, 2004).
\bibitem{Gut05} A.J. Guttmann, Pramana \textbf{64}, 829 (2005).
\bibitem{Fis59} M.E. Fisher, Phys. Rev. {\bf 113}, 969 (1959).
\bibitem{Syo72} I. Syozi, Phase Transition and Critical Phenomena, Vol. 1, edited by C. Domb, and M. S. Green, (Academic Press, New York, 1972), pp. 269--329.
\bibitem{Roj09} O. Rojas, J.S. Valverde, S.M. de Sousa, Physica A {\bf 388}, 1419 (2009).
\bibitem{Str10} J. Stre\v{c}ka, Phys. Lett. A  {\bf 374}, 3718 (2010).
\bibitem{Lis11} B.M. Lisnii, Ukr. J. Phys. {\bf 56}, 1237 (2011).
\bibitem{Str12} J. Stre\v{c}ka, C. Ekiz, Physica A {\bf 391}, 4763 (2012).
\bibitem{Cis13} J. \v{C}is\'arov\'a, J. Stre\v{c}ka, Phys. Rev. B {\bf 87}, 024421 (2013).
\bibitem{Lis14} B. Lisnyi, J. Stre\v{c}ka, Phys. Status Solidi B \textbf{251}, 1083 (2014).
\bibitem{Roj11} O. Rojas, S.M. de Souza, Phys. Lett. A \textbf{375}, 1295 (2011). 
\bibitem{Roj12} M. Rojas, S.M. de Souza, O. Rojas, arXiv:1212.5552.
\bibitem{Per08} M.S.S. Pereira, F.A.B.F de Moura, M.L. Lyra, Phys. Rev. B {\bf 77}, 024402 (2008).
\bibitem{Per09} M.S.S. Pereira, F.A.B.F de Moura, M.L. Lyra, Phys. Rev. B {\bf 79}, 054427 (2009).
\bibitem{Lys11} B.M. Lisnyi, Low Temp. Phys. {\bf 37}, 296 (2011).
\bibitem{Lis13} B.M. Lisnyi, Ukr. J. Phys. {\bf 58}, 195 (2013).
\bibitem{Ana12} O. Rojas, S.M. de Souza, N.S. Ananikian, Phys. Rev. E \textbf{85}, 061123 (2012). 
\bibitem{Nal14} M. Nalbandyan, H. Lazaryan, O. Rojas, S.M. de Souza, N. Ananikian, J. Phys. Soc. Jpn. \textbf{83}, 074001 (2014).
\bibitem{Car14} R.C.P. Carvalho, M.S.S. Pereira, M.L. Lyra, O. Rojas, J. Stre\v{c}ka, Acta Phys. Polonica A \textbf{126}, 12 (2014).
\bibitem{Str09} J. Stre\v{c}ka, A. Tanaka, L. \v{C}anov\'a, T. Verkholyak, Phys. Rev. B {\bf 80}, 174410 (2009).
\bibitem{Tan10} J. Stre\v{c}ka, A. Tanaka, M. Ja\v{s}\v{c}ur, J. Phys.: Conf. Ser. \textbf{200}, 022059 (2010).
\bibitem{Gal11} L. G\'alisov\'a, J. Stre\v{c}ka, A. Tanaka, T. Verkholyak, J. Phys.: Condens. Matter {\bf 23}, 175602 (2011).
\bibitem{Dor14} F.F. Doria, M.S.S. Perreira, M.L. Lyra, J. Magn. Magn. Mater. \textbf{368}, 98 (2014).
\bibitem{Mam99} M. Mambrini, J. Tr\'{e}bosc, F. Mila, Phys. Rev. B {\bf 59}, 13806 (1999).
\bibitem{Roj03} O. Rojas, F.C. Alcaraz, Phys. Rev. B {\bf 67}, 174401 (2003).
\bibitem{Bat03} C.D. Batista, B.S. Shastry, Phys. Rev. Lett. {\bf 91}, 116401 (2003).
\bibitem{Mak11} M. Maksymenko, O. Derzhko, J. Richter, Eur. Phys. J. B {\bf 84}, 397 (2011).
\bibitem{Mak12} M. Maksymenko, O. Derzhko, J. Richter, Acta Phys. Polonica A {\bf 119}, 860 (2011).
\bibitem{Ant09} D. Antonosyan, S. Bellucci, V. Ohanyan, Phys. Rev. B {\bf 79}, 014432 (2009).
\bibitem{Oha10} V. Ohanyan, Phys. Atom. Nucl. {\bf 73}, 494 (2010).
\bibitem{Has08} M. Hase, H. Kitazawa, K. Ozawa, T. Hamasaki, H. Kuroe, T. Sekine, J. Phys. Soc. Jpn. {\bf 77}, 034706 (2008).
\bibitem{Kur10} H. Kuroe, T. Hosaka, S. Hachiuma, T. Sekine, M. Hase, K. Oka, T. Ito, H. Eisaki, M. Fujisawa, S. Okubo, H. Ohta, J. Phys. Soc. Jpn. {\bf 80}, 083705 (2010).
\bibitem{Mat12} M. Matsumoto, H. Kuroe, T. Sekine, M. Hase, J. Phys. Soc. Jpn. {\bf 81}, 024711 (2012).
\bibitem{Kra44} H.A. Kramers, G.H. Wannier, Phys. Rev. {\bf 60}, 252 (1944).
\bibitem{Faz99} P. Fazekas, Lecture Notes on Electron Correlation and Magnetism (Singapore: World Scientific, 1999).
\bibitem{Can06} L. \v{C}anov\'a, J. Stre\v{c}ka, M. Ja\v{s}\v{c}ur, J. Phys.: Condens. Matter {\bf 18}, 4967 (2006).
\bibitem{Man08} F. Mancini, F.P. Mancini, Phys. Rev. E \textbf{77}, 061120 (2008).
\bibitem{Man13} F. Mancini, E. Plekhanov, G. Sica, Eur. Phys. J. B \textbf{86}, 224 (2013). 

\end{thebibliography}
\end{document}